\documentclass[aps,prx,twocolumn,final,letterpaper,10pt]{revtex4-2}

\usepackage{graphicx}% Include figure files
\usepackage{dcolumn}% Align table columns on decimal point
\usepackage{bm}% bold math
\usepackage{amsmath}
\usepackage{xcolor}

\begin{document}

\preprint{APS/123-QED}

\title{Cavity Quantum Electrodynamics in Finite-Bandwidth Squeezed Reservoir}% Force line breaks with \\

\author{Trung Kiên Lê}
\author{Daniil M. Lukin}%

\author{Charles Roques-Carmes}
 
\author{Aviv Karnieli}

\author{Eran Lustig}

\author{Melissa A. Guidry}

\author{Shanhui Fan}

\author{Jelena Vučković}
\affiliation{Ginzton Laboratory, Stanford University, Stanford, CA, United States}

\date{\today}% It is always \today, today,

\begin{abstract}

Light-matter interaction with squeezed vacuum has received much interest for the ability to enhance the native interaction strength between an atom and a photon with a reservoir assumed to have an infinite bandwidth. Here, we study a model of parametrically driven cavity quantum electrodynamics (cavity QED) for enhancing light-matter interaction while subjected to a finite-bandwidth squeezed vacuum drive. Our method is capable of unveiling the effect of relative bandwidth as well as squeezing required to observe the anticipated anti-crossing spectrum and enhanced cooperativity without the ideal squeezed bath assumption. Furthermore, we analyze the practicality of said models when including intrinsic photon loss due to resonators imperfection. With these results, we outline the requirements for experimentally implementing an effectively squeezed bath in solid-state platforms such as InAs quantum dot cavity QED such that \textit{in situ} control and enhancement of light-matter interaction could be realized. 

\end{abstract}

\maketitle

%\tableofcontents

% Need to rewrite introduction 
\section{\label{sec:level1}Introduction}

Squeezed light, in which the uncertainty in one quadrature is reduced below that of the vacuum state, provides powerful resources for various quantum applications such as quantum sensing~\cite{PhysRevX.13.041021}, networking \cite{Kraus2004}, and simulations \cite{Zhu2020}, as well as efficient spin-squeezing \cite{Groszkowski2020} and qubit readouts \cite{Eddins2018, Qin2022, Qin2024}. Recently, squeezed light has attracted a lot of interest as a means to modify and enhance light-matter interaction in cavity QED \cite{Georgiades1995, Turchette1998, Messikh1998, Dalton1999, Zeytinoglu2017, Leroux2018, Qin2018, Chen2019, Li2020, Burd2021, Burd2024, Villiers2024, Qin2024amp, Lukin2024} as well as boson-mediated interaction \cite{Lemonde2016, Burd2019, Arenz2020}. For cavity QED, it has been shown theoretically and experimentally  that using intracavity squeezing combined with a ``squeezed reservoir'' - that is, a quantum reservoir induced by injecting an infinitely broadband squeezed vacuum \cite{Gardiner1986, Murch2013, Toyli2016} - can exponentially improve the interaction strength as well as provide an \textit{in situ} dynamical control of the coupling rate. Coupling the system to the aforementioned ideal squeezed reservoir is necessary for cancelling out the effective thermal noise induced by the squeezed intracavity modes, as was shown in \cite{Leroux2018, Qin2018, Li2020, Qin2024}. 

Squeezing-enhanced cavity QED systems are a promising route for realizing strong and even ultrastrong coupling in systems with otherwise limited cooperativity, which is particularly important for optical cavity QED based technologies. However, understanding the practical performance of such systems can strongly depend on the properties of the squeezed reservoir they are coupled to. In particular, it is known in other contexts that the spectral bandwidth of the injected squeezed light can affect the properties of cascaded quantum optical systems \cite{Crisafulli2013, Zippili2014, Asjad2016, Gross2022, Joshi2023}. Unfortunately, existing theoretical frameworks to date fail to model such inter-cavity, corresponding to finite squeezing bandwidths. Instead, a simplifying assumption of infinite bandwidth (i.e., one large enough compared to any other spectral scale in the system) allows accurate application of the Born-Markov approximation to eliminate the reservoir's self-correlations \cite{Murch2013}. This assumption, however, is idealized, and it is not clear whether realistic, finite-bandwidth squeezed reservoirs can satisfy the requirements for squeezing-enhanced cavity QED models.

 % for arbitrarily strong squeezing strength. 

In this work, we present a general framework for exploring squeezing-enhanced cavity QED systems with finite-bandwidth squeezed reservoirs. We demonstrate how the broadband squeezed reservoir assumption breaks down for various cavity QED effects at varying bandwidths, particularly in the cases of enhanced light-matter interactions and dynamics of the qubit and cavity under finite-bandwidth squeezed reservoirs. We show that for sufficiently large bandwidth, the exponential enhancement between the Bogoliubov mode and atom is recovered, but for bandwidths below a certain point, anti-crossing spectra are modified due to effective thermal heating of the squeezed vacuum state and no strong-coupling splitting is observed. Furthermore, we find that including intrinsic loss in the cascaded system has a significant impact on the effectiveness of enhancing light-matter interaction via intracavity squeezing (ICS) and injected external squeezed (IES) light. Our findings pave the way towards understanding the interplay between squeezing bandwidth and the effectiveness of injected squeezed vacuum in suppressing unwanted decoherences.

\section{Quantum system in a squeezed reservoir}

\begin{figure*}[t!]
    \centering
    \includegraphics[width = 18cm]{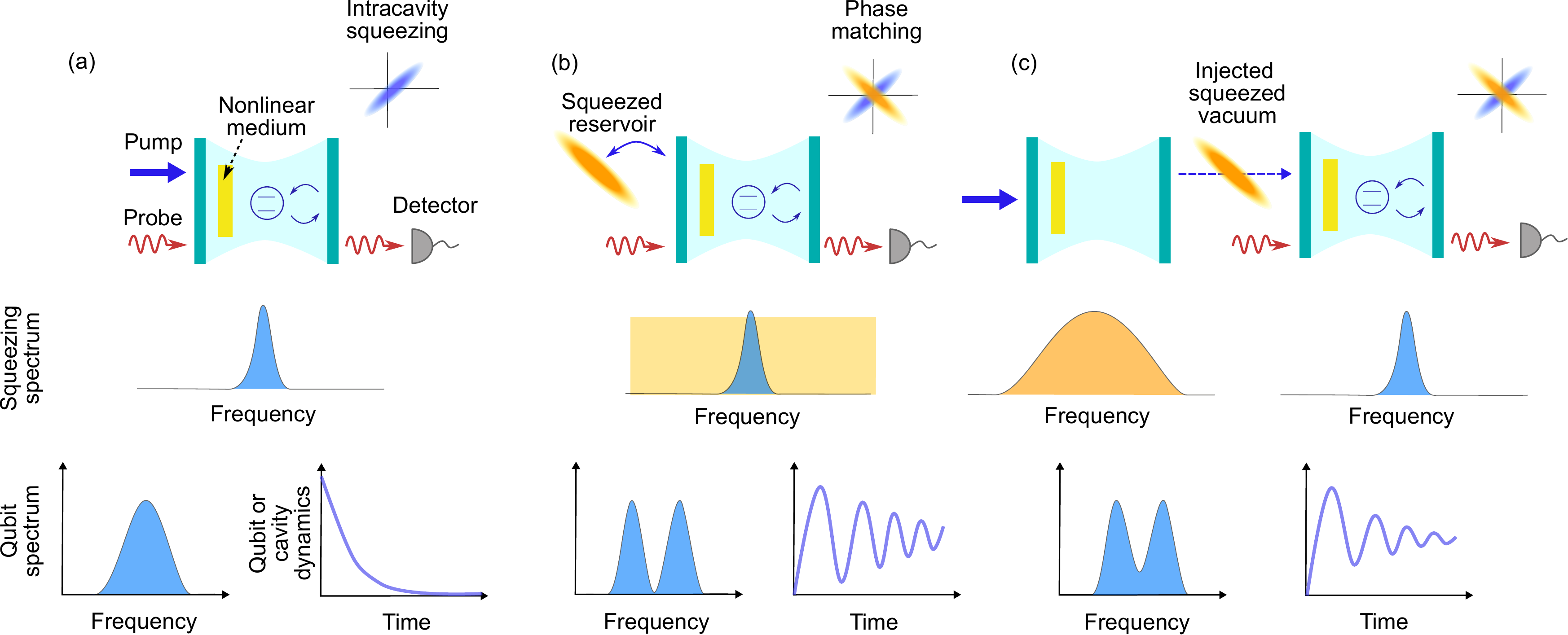}
    \caption{\textit{Illustration:} \textbf{(a)} A cavity QED system with the cavity (cyan) coupled to a two-level atom (center circle) is driven by a degenerate parametric process, represented by pump and a nonlinear medium (yellow), and the probe's photons scattering from the systems are detected to measure either spectrum or time dynamics. The cavity itself does not couple to a squeezed bath, resulting in irreversible decay in case of the weakly coupled cavity QED system. \textbf{(b)} The idealized case where a cavity QED system with squeezing is coupled to a perfect squeezed bath (orange ellipse) whose spectrum is a white noise spectrum (i.e., an infinite bandwidth). In this scenario, exponential enhancement of light-matter interaction due to squeezing exhibits strong coupling spectrum and observable Rabi dynamics of the qubit and cavity, even when the qubit itself is weakly coupled to the cavity mode, as in (a). \textbf{(c)} Injected large-bandwidth squeezed vacuum, where the idealized squeezed reservoir is replaced with a large squeezed bandwidth source, and whose output is then injected into a cavity QED system with squeezing. This protocol still enhances the light-matter interaction, but still has residual unwanted decoherence that makes its spectrum and Rabi dynamics less visible than that of the perfect squeezed reservoir case.}
    \label{fig:SLH_fig}
\end{figure*}

% Maybe put figure 1 here to illustrate the SLH as an intro figure 

% Advantages and disadvantages 

% Need to decide where to add figures in here 

\subsection{Cavity QED in a squeezed resonator}
We first rederive the master equation of a parametrically-driven cavity QED system coupled to a broadband squeezed reservoir. Our cavity system comprises a detuned degenerate optical parametric oscillator (DOPO) with detuning $\delta_a$ and parametric drive strength $E_a$ (assumed to be real value), and can be captured by the following Hamiltonian 
\begin{equation}
    H_\text{DOPO} = \delta_a \hat{a}^{\dagger} \hat{a} - E_a (\hat{a}^{\dagger 2} + \hat{a}).
\end{equation}
We note that the detuned DOPO is diagonalizable by the Bogoliubov transform and is dynamically stable without dissipation \cite{Metelmann2022} as long as $\delta_a > E_a$ \cite{Collet1984}. Our DOPO cavity also contains a two-level atom (emitter), i.e., qubit, whose coupling to the cavity is described by the Jaynes-Cumming model. Thus, the total cavity Hamiltonian becomes 
\begin{equation}
    H = \delta_a \hat{a}^\dag \hat{a} + \frac{\delta_q}{2} \hat{\sigma}_z - E_a (\hat{a}^{\dagger 2} + \hat{a}^2) + g( \hat{a}^\dag \hat{\sigma} + \hat{a} \hat{\sigma}^\dag ).
\end{equation}
We now introduce a Bogoliubov transformation on the cavity operator \cite{Munoz2020}
\begin{align}
    \hat{\alpha} = \hat{a} \cosh r + \hat{a}^\dag \sinh r,
\end{align}
where $r$ is the squeezing coefficient defined by 
\begin{align}
    r = \frac{1}{4} \ln \frac{1 + \lambda}{1 - \lambda},
\end{align}
and $\lambda = E_a / \delta_a$. This transformation diagonalizes the DOPO and defines a Bogoliubov oscillator (BO) in terms of the new ladder operator $\hat{\alpha}$ with a renormalized frequency. One then obtains a new Hamiltonian in the Bogoliubov frame 
\begin{align}
    \Tilde{H} = \frac{\delta_a}{\cosh^2 r}\hat{\alpha}^\dag \hat{\alpha} + \frac{\delta_q}{2} \hat{\sigma}_z + \frac{ge^r}{2} (\hat{\alpha}^\dag + \hat{\alpha}) \hat{\sigma}_x \notag \\ - i\frac{ge^{-r}}{2} (\hat{\alpha}^\dag - \hat{\alpha}) \hat{\sigma}_y
\end{align}

From this Hamiltonian, one can see that in the Bogoliubov frame, the coupling strength between an atom and a bosonic mode grows exponentially with $r$ as $ge^r/2$. However, this picture is drastically changed once loss is included, as the cavity loss operator is transformed accordingly into the squeezed frame as~\cite{Shani2022}:
\begin{align}
    L_{\hat{\alpha}} = \sqrt{\kappa} (\hat{\alpha} \cosh r + \hat{\alpha}^\dag \sinh r).
\end{align}

Under the action of the squeezed frame Lindbladian, decoherence is also exponentially amplified, which can wash away vacuum Rabi splitting in either the qubit or the cavity's output spectrum, as illustrated in Fig. (\ref{fig:SLH_fig}a). It has been suggested that coupling a cavity QED system to a broadband squeezed vacuum reservoir in terms of the new ladder operator $\hat{\alpha}$ with a renormalized frequency, where the squeezing bandwidth is vastly larger than the cavity's linewidth - as shown in Fig. (\ref{fig:SLH_fig}b), could mitigate or even cancel out the effect of $L_{\hat{\alpha}}$ \cite{Leroux2018, Qin2018, Munoz2020}. Working in this infinite-bandwidth squeezed reservoir approximation, we find that the Lindbladian master equation takes the form  
\begin{align}
    \frac{d\rho}{dt} = &-i[H, \rho] + \frac{\kappa}{2}(N + 1 + \eta)D_a [\rho] + \frac{\kappa}{2} N D_{a^\dagger} [\rho] \notag \\ &- \frac{\kappa}{2} M D'_a [\rho] - \frac{\kappa}{2} M^* D'_{a^\dagger} [\rho],
\end{align}
where $D_a[\rho] = a \rho a^\dagger - \frac{1}{2} \{a^\dagger a, \rho \}$ and $D'_a[\rho] = a \rho a - \frac{1}{2} \{a a, \rho \}$. Throughout this paper, we use $D_a[\rho]$ and $D[a](\rho)$ interchangeably. $N, M$ are the effective thermal photon and two-photon coefficients introduced by the squeezed bath, respectively, and are defined as 
$N = \sinh^2 r_{\mathrm{e}}, M = e^{i \theta_e} \sinh r_{\mathrm{e}} \cosh r_{\mathrm{e}} $, where $r_{\mathrm{e}}$ denotes the reservoir's squeezing strength and $\theta_e$ its quadrature phase. Above, $\eta$ is the ratio between the cavity's intrinsic loss rate $\kappa_i$ and outcoupling rate $\kappa$, i.e. $\eta = \kappa_i / \kappa$. More compactly (and, for the discussion in this subsection, assuming $\eta=0$), we can represent the Lindbladian with a single dissipator as %To write this equation in a way compatible with QuTiP
\begin{align}
    \frac{d\rho}{dt} = -i[H, \rho] + \kappa D_{\sqrt{N + 1} \hat{a} - e^{i\theta} \sqrt{N} \hat{a}^\dagger} [\rho].
    \label{eq:sqz_bath}
\end{align}
With the right squeezing strength and phase-matching conditions, which ensure that the intra-cavity squeezed vacuum transforms to a vacuum state in the Bogoliubov frame, (see Appendix B for derivation), we can write Eq. (\ref{eq:sqz_bath}) in the Bogoliubov frame as
\begin{align}
     \frac{d\tilde{\rho}}{dt} = -i[\Tilde{H}, \tilde{\rho}] + \kappa D_{\hat{\alpha}} [\tilde{\rho}].
     \label{eq:sqz_bath_bogo}
\end{align}
In Eq. (\ref{eq:sqz_bath_bogo}), the Bogoliubov-frame state $\tilde{\rho}$ is now initialized to a vacuum state owing to the phase matching condition. Moreover, the loss rate stays the same as in the lab frame while the coupling is exponentially enhanced, resulting in true exponential enhancement in cooperativity for the atom-photon interaction. Intuitively, this effect can be understood as an exponential amount of photons coherently interacting with the atom inside the cavity, while at the same time the squeezed reservoir counteracts the associated noise induced by the intracavity squeezed state, thereby suppressing unwanted decoherences. 

\subsection{Finite-bandwidth squeezed bath}

While the broadband squeezed reservoir assumption simplifies the analysis for squeezing-enhanced cavity QED systems, it has several unphysical properties, such as requiring infinite energy \cite{Gross2022}. Furthermore, it remains rather vague what are the bandwidth regimes for which the broadband squeezed reservoir assumptions can hold in experimental conditions, and at arbitrarily strong squeezing coefficient of the drive. In this section, we derive the requisite master equation to capture the effects of finite-bandwidth squeezed vacuum injection. 

%In our model, the broadband squeezed vacuum reservoir is replaced by a finite-bandwidth degenerate optical parametric oscillator (OPO) source, whose output is then cascaded into a quantum system of interest \cite{Gardiner1994, Georgiades1995, Dalton1999, Turchette1998}. Modelling such cascaded quantum system can be done using the SLH formalism \cite{Combes2017} that models cascades as coupling between different quantum systems, which has also been used to capture chiral quantum optical system \cite{Joshi2023} and quantum feedback \cite{Crisafulli2013}. We emphasize that while driving squeezed vacuum into a quantum system of interest has been studied theoretically and experimentally \cite{ Georgiades1995, Murch2013, Toyli2016, Eddins2018}, there is to this date no systematic study of how the injected squeezed vacuum's bandwidth affects the dynamics and interaction properties of cavity QED systems.

When relaxing the infinite bandwidth approximation, the master equation no longer obeys a pure Lindbladian or Markovian evolution as the system-bath exhibits non-zero temporal correlations (or memory) of the form $\langle a^\dag_i(t) a_j(t') \rangle$ \cite{Zippili2014, Asjad2016, Gross2022}, making it rather difficult to derive a rigorous, self-consistent non-Markovian master equation, especially when the system contains both photonic and atomic degree of freedom. A more accurate model that can account for finite bandwidth whilst allowing a relatively simple master equation can be derived from the cascaded system via the SLH formalism \cite{Combes2017}, as illustrated in Fig. \ref{fig:SLH_fig}(c). For our system of interest, we write the jump operators as $L_1 = \sqrt{\kappa_a} \hat{a} \textrm{ and } L_2 = \sqrt{\kappa_b} \hat{b}$, with additional intrinsic loss operators $L_{1,i} = \sqrt{\kappa_i} \hat{a}, L_{2,i} = \sqrt{\kappa_i} \hat{b}$. The bandwidth of our ``squeezed reservoir'' in SLH is determined by the relative magnitude of $\kappa_a \textrm{ to } \kappa_b$, which are the coupling rates of cavity modes $\hat{a}, \hat{b}$, respectively, into the common waveguide. We then get the full system's master equation as:
\begin{align}
    \frac{d \rho}{dt} &= - i[H, \rho] + D[\sqrt{\kappa_a} \hat{a} + \sqrt{\kappa_b} \hat{b}](\rho) \notag \\ &+ D[\sqrt{\kappa_i} \hat{a}] (\rho) + D[\sqrt{\kappa_i} \hat{b}] (\rho) \\ %D[\sqrt{\kappa_a + \kappa_i} \hat{a} + \sqrt{\kappa_b + \kappa_i} \hat{b}](\rho)\\
    H &= i(E_a \hat{a}^{\dagger 2} - E_a^* \hat{a}^2) + \delta_b \hat{b}^\dagger \hat{b} + (E_b \hat{b}^{\dagger 2} + E_b^* \hat{b}^2) \notag \\ & + \frac{\sqrt{\kappa_a \kappa_b}}{2i}(\hat{b}^\dagger \hat{a} - \hat{a}^\dagger \hat{b} )
    \label{eq:SLH_OPO}
\end{align}
Where $E_b$ is the parametric drive amplitude of the cavity $\hat{b}$. Note that we choose the first degenerate OPO to be on-resonant instead of detuned: in this manner, it reduces to an ideal squeezing source such that the output light is perfectly squeezed in the absence of intrinsic loss. We note that in the limit of $\kappa_a \rightarrow \infty$, the broadband squeezed reservoir approximation is recovered via adiabatic elimination of the $\hat{a}$ mode (see Appendix C for details). 

\subsection{Cavity QED in finite-bandwidth squeezed bath}

With the SLH formalism, we can straightforwardly extend our model to include a cavity QED system with both injected and intracavity squeezing. By coupling a qubit with a spontaneous decay process $ D[\sqrt{\gamma} \sigma]$ to mode $\hat{b}$ and performing the Bogoliubov transformation on the $\hat{b}$ as $ \hat{\beta} = \hat{b} \cosh r - \hat{b}^\dagger \sinh r $, we obtain the Bogoliubov frame Hamiltonian of the cascaded model, where the atom is now coupled to the Bogoliubov mode $\hat{\beta}$
% \begin{widetext}
%     \begin{align}
%         \tilde{H} = i(E_a \hat{a}^{\dagger 2} - E_b^* \hat{a}^2) + \frac{\delta_b}{\cosh^2r} \hat{\beta}^\dagger \hat{\beta} + \frac{g e^r}{2}(\hat{\beta} + \hat{\beta}^\dagger)(\sigma + \sigma^\dagger) + \frac{\sqrt{\kappa_a \kappa_b}}{2i}\left[ (\hat{\beta}^\dagger \cosh r + \hat{\beta} \sinh r) \hat{a} - \textrm{H.c} \right] + H_\text{err} \\
%          H_\text{err} = - \frac{g e^{-r}}{2}(\hat{\beta} - \hat{\beta}^\dagger)(\sigma - \sigma^\dagger) \\
%         \frac{d \tilde{\rho}}{dt} = - i[\tilde{H}, \tilde{\rho}] + D[\sqrt{\kappa_a} \hat{a} + \sqrt{\kappa_b} ( \hat{\beta} \cosh r + \hat{\beta}^\dagger \sinh r ) ](\tilde{\rho}) + D[\sqrt{\kappa_i} \hat{a}] (\tilde{\rho}) + D[\sqrt{\kappa_i} \hat{b}] (\tilde{\rho}) + D[\sqrt{\gamma} \sigma](\tilde{\rho})  \label{eq:SLH_cavqed} 
% \end{align}
% \end{widetext}

\begin{figure*}[t!]
    \centering
    \includegraphics[width = \textwidth]{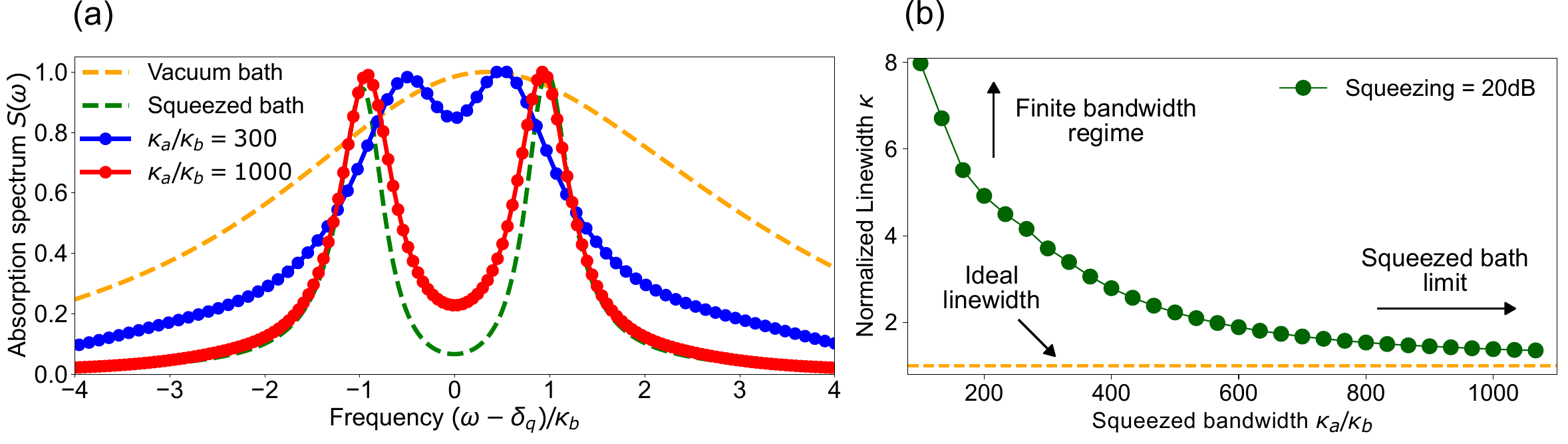}
    \caption{\textit{Weakly coupled system}. We solve the master equation for spectra using  $\delta_\beta = \delta_q = 1.0, \kappa_a \in [1.5, 16.5], E_a = 0.2055 \kappa_a, \kappa_b = 0.015, \gamma = 0.001$ and $g = 0.003$. All units are normalized to $\delta_q$. \textbf{(a)} For a system fixed at $20$dB squeezing, we vary the bandwidth ratio $\kappa_a/\kappa_b$. Using parameters for a weakly coupled system, we plotted the atom's absorption spectrum and compare against the spectrum obtained from squeezed bath as well as a vacuum bath. For $\kappa_a/\kappa_b = 300$, the spectrum has a large linewidth, indicating that the system is experiencing squeezing-induced effective thermal decoherence process. For $\kappa_a/\kappa_b = 1000$, the spectrum is comparable to the ideal squeezed bath case (green dashed line). \textbf{(b)} Convergence towards squeezed bath can be achieved through increasing the required bandwidth $\kappa_a$, while at finite-bandwidth regime the absorption spectrum is dominated by thermal and squeezing noises.}  % need to take log-log plot 
    \label{fig:Weak_VRS}
\end{figure*}

\begin{widetext}
    \begin{align}
        \tilde{H} & = i(E_a \hat{a}^{\dagger 2} - E_b^* \hat{a}^2) + \frac{\delta_b}{\cosh^2 r} \hat{\beta}^\dagger \hat{\beta} + \frac{g e^r}{2}(\hat{\beta} + \hat{\beta}^\dagger)(\sigma + \sigma^\dagger) + \frac{\sqrt{\kappa_a \kappa_b}}{2i}\left[ (\hat{\beta}^\dagger \cosh r + \hat{\beta} \sinh r) \hat{a} - \textrm{H.c} \right] + H_\text{err} \label{eq:SLH_cavqed} \\
        H_\text{err} & = - \frac{g e^{-r}}{2}(\hat{\beta} - \hat{\beta}^\dagger)(\sigma - \sigma^\dagger)   
    \end{align}
\end{widetext}
And the master equation can be similarly obtained as 
\begin{widetext}
\begin{align}
        \frac{d \tilde{\rho}}{dt} & = - i[\tilde{H}, \tilde{\rho}] + D\left[\sqrt{\kappa_a} \hat{a} + \sqrt{\kappa_b} \left( \hat{\beta} \cosh r + \hat{\beta}^\dagger \sinh r \right) \right](\tilde{\rho}) + D\left[\sqrt{\kappa_i} \hat{a}\right](\tilde{\rho}) + D\left[\sqrt{\kappa_i} \hat{b}\right](\tilde{\rho}) + D\left[\sqrt{\gamma} \sigma\right](\tilde{\rho}) \label{eq:SLH_cavqed_me}
\end{align}
\end{widetext}

Here, we use $D[\sqrt{\kappa_i} \hat{b}] (\tilde{\rho})$ to indicate intrinsic loss channel with $\hat{b} = \hat{\beta} \cosh r + \hat{\beta}^\dagger \sinh r$. In the Bogoliubov frame $\hat{\beta}$, we have the simultaneous exponential enhancement to the atom-cavity coupling rate $g$ as $g_s = ge^r / 2$ to the BO \textit{and} dissipation rate $\kappa_b$. This enhanced decoherence has been observed in superconducting qubit \cite{Villiers2024} and trapped ion experiments \cite{Burd2021, Burd2024}, and analyzed theoretically in \cite{Lemonde2016, Shani2022}, where the squeezing exponentially amplifies all bosonic decoherence channels such as thermal and dephasing processes. In the Bogolibov frame, these decoherence channels manifest as effective thermal noise that scales as $\sinh^2 r$ and two-photon noises that scales as $\cosh r \sinh r$ \cite{Villiers2024}. By replacing the squeezed bath with a source OPO that injects squeezed light into $\hat{b}$, we can explore the necessary bandwidth $\kappa_a / \kappa_b$ to effectively cancel out said noises and observe the enhanced coupling $g_s$, which can be extracted from the qubit absorption spectrum \cite{Leroux2018, Wallraff2004}
\begin{align}
    S(\omega) = \int_{-\infty}^{\infty} dt e^{-i\omega t} \langle \hat{\sigma}_-(t) \hat{\sigma}_+(0) \rangle 
    \label{eq:AbsSpec},
\end{align} 
where $\hat{\sigma}_+(0)$ is the steady-state expectation value of $\hat{\sigma}_+$, defined as $\hat{\sigma}_+(0) \equiv \lim_{t \rightarrow \infty}\textrm{Tr}[\rho(t) \hat{\sigma}_+]$. The use of qubit absorption spectrum allows a true lab frame interpretation of light-matter interaction, and extraction of the enhanced coupling as the difference between the frequencies of the two peaks. 

\begin{figure*}[t!]
    \centering
    \includegraphics[width = \textwidth]{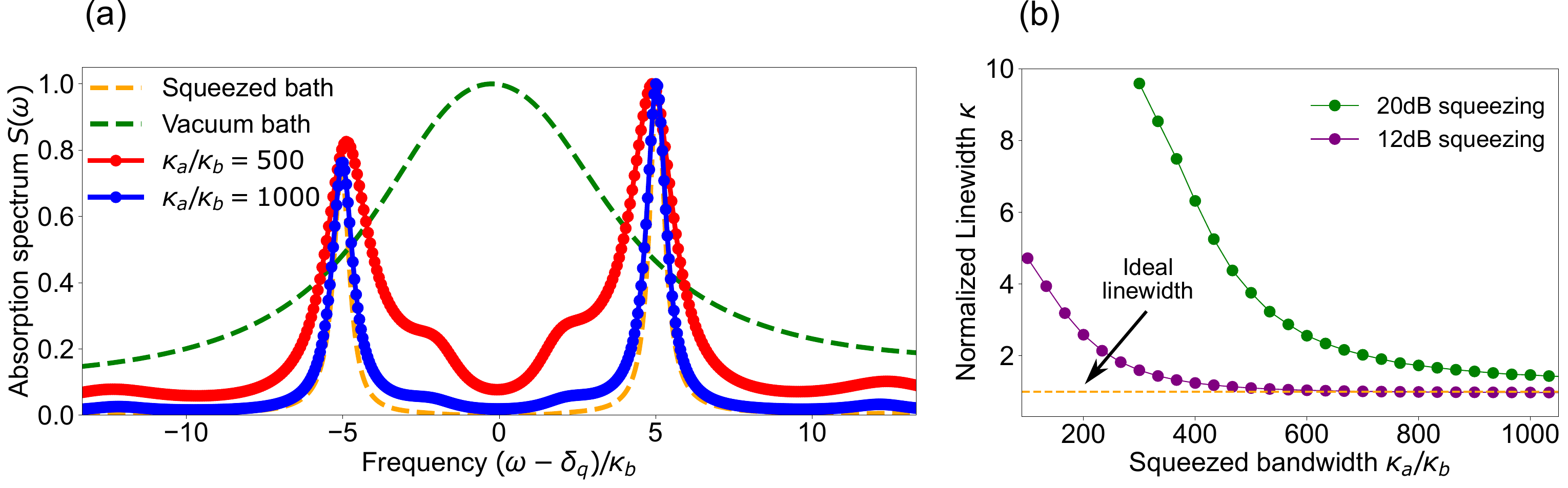}
    \caption{\textit{Strongly coupled system:} We set $\delta_\beta = \delta_q = 1.0, \kappa_a \in [3.0, 16.5], E_a = 0.2055 \kappa_a, g = \kappa_b = 0.015$ and $\gamma = 0.001$ for our master equation simulation. All units are normalized to $\delta_q$. \textbf{(a)} For a strongly coupled system, we plotted the anti-crossing spectra and compare against the absorption spectrum obtained from squeezed bath as well as a vacuum bath. For $\kappa_a/\kappa_b = 500$, the spectrum has a large linewidth and signature of thermal noise as shown by extra smaller peaks, and for $\kappa_a/\kappa_b > 1000$, the spectrum is comparable to the ideal squeezed bath case (orange dashed line) with minimal residual thermal noises (shown by the near-absence of smaller extra peaks). \textbf{(b)} Comparison of spectra linewidths for 12dB and 20dB squeezing, showing the relatively fast convergence towards the ideal squeezed bath limit for 12dB, indicating the dependence of the squeezed bath assumption's validity on the squeezing strength.} % need to take log-log plot 
    \label{fig:strong_coupling}
\end{figure*}

\section{Results}

% Also, fit linewidth versus ka/kb 

% We performed simulations using Eq. (\ref{eq:SLH_OPO}) to quantify deviations for the case of an OPO cavity coupled to broadband squeezed reservoir and a cascaded OPO system where both OPOs have finite bandwidth. The observable we used here is $\langle 0_s | \beta_s^\dagger \beta_s | 0_s \rangle \equiv N_s$, the number of squeezed photon in the Bogoliubov basis. 

\subsection{Enhancing weakly coupled cavity QED}

The simplest application of our cascaded model is to enhance a weakly-coupled system \cite{Leroux2018}. The finite but large-bandwidth squeezed vacuum effectively ``cools'' the BO down to its ground state, which in the lab frame is equivalent to the squeezed vacuum (See Appendix D for details). Here, we show that through numerically solving Eq.~(\ref{eq:SLH_cavqed_me}) and Eq.~(\ref{eq:AbsSpec}) we can find exactly how much bandwidth --- set by the ratio $\kappa_a / \kappa_b$ --- one needs to practically recover the exponential enhancement in coupling strength, which we compare against the spectra obtained from eq. (\ref{eq:sqz_bath_bogo}). We first set $\kappa_i = 0$ to study a simplified model without intrinsic loss, and for $g/\kappa_b < 0.5$ or cooperativity $C = 4g^2 / \kappa_b \gamma < 1$, the system is in the weak-coupling regime of cavity QED. The choice of $E_a$ gives a reservoir steady-state squeezing of $\approx 20$dB, similar to the squeezing value reported in \cite{Leroux2018}. 

% Omit data below ka/kb = 450 for 20dB squeezing but for 12dB keep it 
% Also, fit linewidth versus ka/kb 

In Fig. \ref{fig:Weak_VRS}a, we show the absorption spectrum for different bandwidths with a fixed squeezing of $20$dB. When the source bandwidth goes from $\kappa_a / \kappa_b = 100$ to $\kappa_a / \kappa_b > 1000$, vacuum Rabi splitting becomes more pronounced, and eventually converges towards the spectrum of an exponentially enhanced BO-atom system in squeezed bath. We notice that at $\kappa_a / \kappa_b = 300$, the spectrum begins to exhibit the splitting signature of dressed Jaynes-Cumming eigenstates, in spite of excessive loss and linewidth broadening. When the bandwidth is increased, a ``cooling'' effect then starts to kick in, due to suppression of the effective thermal noise $\sinh^2 r$, and can be observed in the form of a narrower linewidth for the value $\kappa_a / \kappa_b = 1070$, as compared to the linewidth at $\kappa_a / \kappa_b = 300$. This confirms that injecting an extremely broadband squeezed vacuum - generated either through a large out-coupling rate $\kappa_a$ or by another mechanism - is equivalent to modelling the bosonic Lindblad operator as an infinite bandwidth squeezed reservoir, with the squeezed thermal noise completely suppressed at the infinite bandwidth limit. 

Motivated by the observation of linewidth narrowing arising from increasing the bandwidth of the squeezed vacuum drive, we plotted the spectrum's linewidth, normalized to the ideal linewidth obtained from computing the spectrum in squeezed bath in Fig. \ref{fig:Weak_VRS}b as a function of $\kappa_a/ \kappa_b$ for different squeezing strength. We observe that the linewidth rapidly decreases as a function of $\kappa_a / \kappa_b$ and eventually converges towards the expected linewidth of a system with given $\kappa_b, \gamma$. 
We note that the required bandwidth for full suppression of the unwanted squeezing-induced effective thermal noise in all of our simulations scales with the squeezing coefficient, and hence the final synthetic coupling strength $ge^r / 2$. These results indicate possibly stringent requirements on pushing a weakly coupled system into a strongly coupled one through simultaneous injected squeezed vacuum with intracavity squeezing. However, a possible application of such a weakly coupled system being enhanced towards strong coupling is squeezed lasing, where the main requirement is having an enhanced cooperativity greater than 1 \cite{Munoz2020}.

\subsection{Effects on strongly coupled system}

\begin{figure*}[t!]
    \centering
    \includegraphics[width = 17cm]{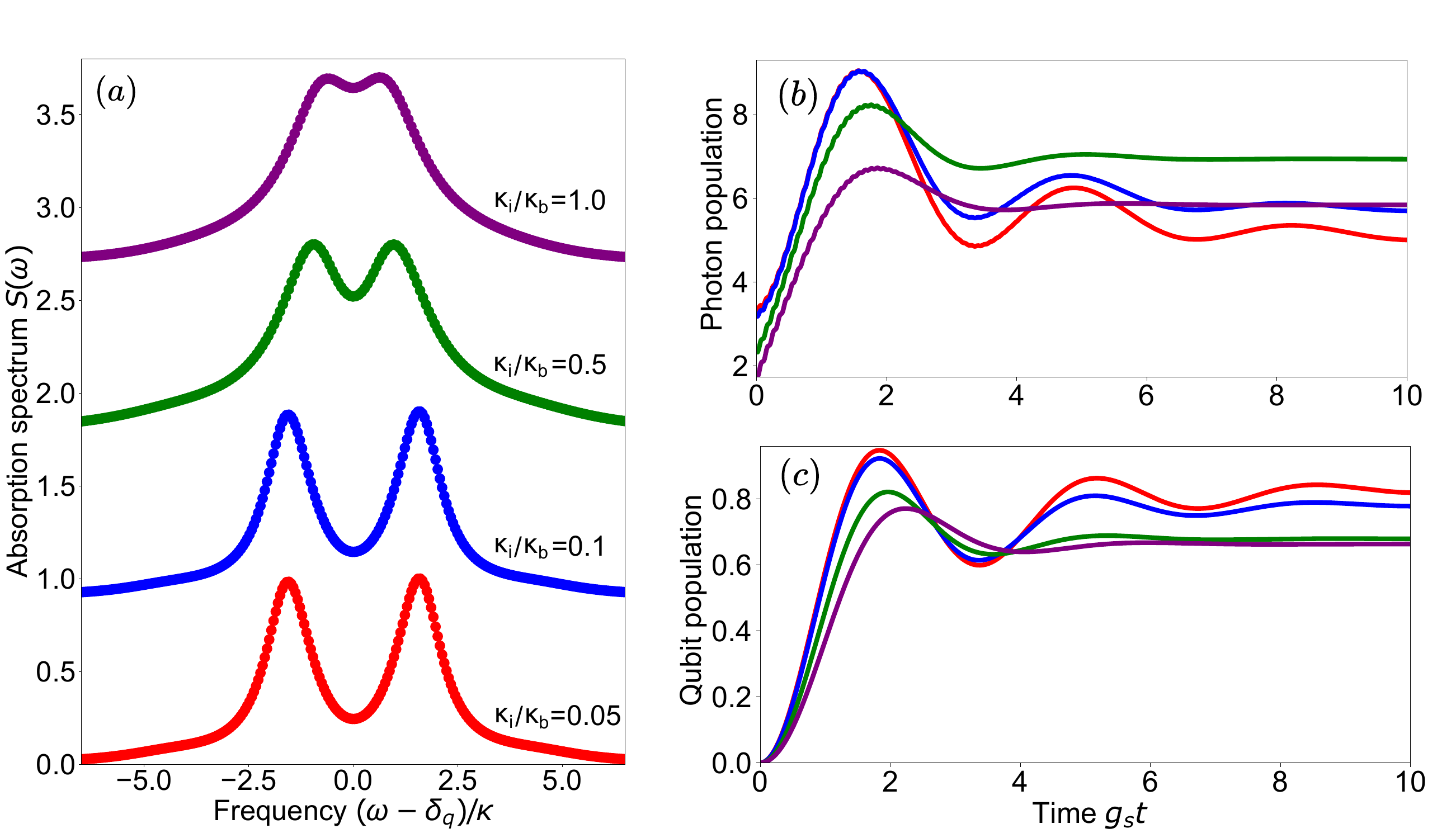}
    \caption{\textit{Effects of intrinsic loss:} \textbf{(a)} Absorption spectra with intrinsic loss $\kappa_i$, where $\kappa_i$ is relative to $\kappa_b$, the outcoupling rate of $\hat{b}$, at 12dB squeezing. For a strongly over-coupled resonator where $\kappa_i / \kappa_b = 0.1$, the anti-crossing spectrum does not change much from the case without intrinsic loss, indicating the negligible squeezed thermal photon contribution, but as soon as $\kappa_i \geq 0.5 \kappa_b $, the spectrum broadens and limits the broadband squeezed vacuum drive, as well as exhibiting features indicative of thermal noises. \textbf{(b)} Vacuum Rabi oscillation of the cavity occupation in the lab frame, \textbf{(c)} and of the qubit. The increase of thermal squeezed photons at higher $\kappa_i$ quickly washes away Rabi oscillation, as shown in the purple data line.}
    \label{fig:intrinsic_loss}
\end{figure*}

Beyond enhancing a weakly-coupled atom-cavity system, the above scheme can also be applied towards enhancing a system already within the strong coupling regime, where $g > \kappa/2 \gg \gamma$. %In this section, we set $\delta_\beta = \delta_q = 1.0, \kappa_a \in [3.0, 16.5], E_a = 0.2055 \kappa_a, g = \kappa_b = 0.015$ and $\gamma = 0.001$, which puts the system in the strong coupling regime. 
Owing to the relative frequency of the Bogoliubov oscillator and the enhanced coupling strength $g_{\textrm{S}} \approx 0.075$, the system exhibits asymmetric spectra that deviates from the Jaynes-Cumming model's anti-crossing response, which can be explained through a Jaynes-Cumming model with re-normalized coupling strength \cite{Cao2011}. 

% Add supplementary figures to explain these thermal peaks for Appendix D 
While a weakly coupled system enhanced towards strong coupling exhibits the standard two-peak vacuum Rabi splitting in its spectra, the enhanced strongly coupled system exhibits multi-peak spectra, where small peaks between and outside the main Rabi splitting peaks arise from multiple higher rung transitions in the atom-cavity system \cite{Tian1992, Wallraff2004} that result from excitations of the higher energy levels of the hybrid cavity-atom system. Specifically, when the squeezing bandwidth is increased, a suppression can be observed via reduction in visibility of multiple smaller peaks exhibited by the spectra, as effective thermal effect - and thereby, higher rung transitions induced by said thermal bath - is reduced. Due to the simultaneous appearance of higher-rung transition peaks and vacuum Rabi splitting for $200 < \kappa_a / \kappa_b < 450$, we choose not to estimate linewidths from these spectra in Fig. (\ref{fig:strong_coupling}b). We note that these thermal-induced transitions start to be suppressed when $\kappa_a / \kappa_b \geq 450$ and become negligible around $\kappa_a / \kappa_b > 800$, once again indicating that the system is "cooling" down to true squeezed vacuum and converges towards the ideal squeezed bath limit as the bandwidth becomes extremely large. For detailed discussions on higher rung transition peaks and noises in the finite-bandwidth SLH "squeezed reservoir", see Appendix D.

Within the strong coupling regime, we explore the interplay between bandwidth and squeezing strength, which provides insights on the validity of the infinite-bandwdith squeezed bath assumption. Interestingly, we observe a trade-off between linewidth and squeezing, which is related to the coherence time of each squeezing quadrature \cite{Gardiner1994}. Intuitively, at smaller squeezing values, the effective squeezed bandwidth of the OPO is increased, which allows reaching closer to the ideal squeezed bath limit even for more modest values of $\kappa_a / \kappa_b$. In Fig. \ref{fig:strong_coupling}b, we observe said trade-off for the case with 20dB and 12dB squeezed vacuum ($r \approx 1.38$), indicating that at 12dB the system converges to the squeezed-bath spectra around $\kappa_a / \kappa_b \gtrsim 400$. Furthermore, at 12dB squeezing, the effective residual squeezed thermal photon occupation is significantly less than that of 20dB squeezing, which means the system is subjected to less thermal noises and therefore a moderately-large-bandwidth squeezed vacuum injection can act effectively as a true squeezed bath. We conclude that the validity of squeezed bath assumption strongly depends on the squeezing coefficient $r$, in addition to the squeezing bandwidth. 

%Photonic crystal cavity SiC device, with the broadband squeezed source is a $\chi^{(2)}$ PPLN waveguide pumped at second-harmonic and generates output squeezed vacuum at $\omega_p / 2$. The color center (circle with pointing arrow) acts as the qubit. The top waveguide couples another two-tone pump with a detuning $\delta$ from the cavity's resonance.  \textbf{(b)} Similar scheme using a pump at second-harmonic instead.

% Fix this with 12dB simulation + squeezing compensation 
\subsection{Intrinsic loss}

\begin{figure}[t!]
    \centering
    \includegraphics[width = 9cm]{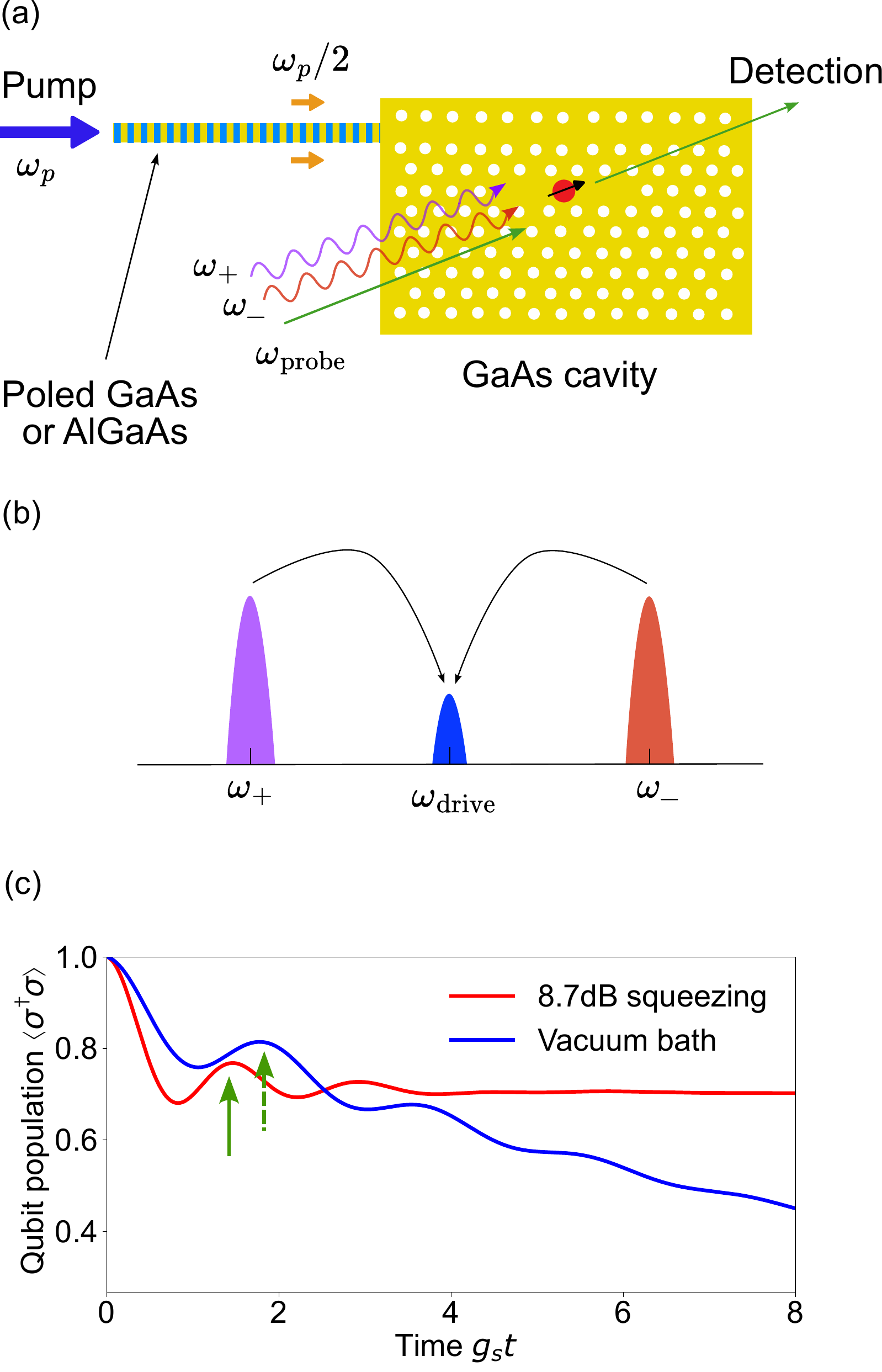}
    \caption{\textit{Experimental proposal}: \textbf{(a)} GaAs photonic crystal cavity with InAs quantum dot (red circle), with poled GaAs or AlGaAs waveguide for facilitating quasi-phase matching for $\chi^{(2)}$ and degenerate squeezing. $\omega_+, \omega_-$ are the pump frequencies for generating  squeezed drive $\omega_\textrm{drive}$ in the cavity using $\chi^{(3)}$ and $\omega_\textrm{probe}$ is the excitation frequency for the quantum dot. Excitation, nonlinear optics and detection of photonic crystal cavity can be performed confocally. 
    \textbf{(b)} Illustration of parametric $\chi^{(3)}$ process to generate degenerate squeezing in a cavity, with the phase matching condition $\omega_+ + \omega_- = 2 \omega_\textrm{drive}$.
    \textbf{(c)} Simulated qubit's Rabi oscillation signature for squeezing on and squeezing off, and in both cases the system is subjected to $10\%$ intrinsic loss and the qubit starts in the excited state $|e\rangle$. Solid (dashed) arrow indicates the Rabi cycle's peak of the squeezed cavity QED's qubit dynamics (vacuum cavity QED's qubit dynamics).}
    \label{fig:proposal}
\end{figure}

The cascaded model can be extended to capture loss by adding an additional decay rate $\kappa_i$ to the Lindbladian. This intrinsic loss is important for capturing the loss of squeezing through photons escaping the cavity into free space, absorption, or other dissipation channels, and is a central issue in efficiently outcoupling or transporting squeezed light generated in a cavity. In Fig. \ref{fig:intrinsic_loss}a we show how intrinsic loss in transporting squeezed light can become a significant impediment to implementing squeezed drive with intracavity squeezing for enhancing cavity QED. 
As expected, a small amount of intrinsic loss ($\kappa_i / \kappa_b = 0.05, 0.1$) does not adversely broadens the spectrum's linewidth nor introduces excessive thermal noise, and only moderately reduces the achievable squeezing/anti-squeezing as seen by the $\hat{b}$ mode. At the value of $\kappa_i / \kappa_b = 1.0$, however, intrinsic loss is sufficient to wash away Rabi oscillations. To understand this, we can see that in the Bogoliubov frame, thermal noises induced by increasing intrinsic loss is "enhanced" by anti-squeezing, which becomes another contribution to the broadening of the anticrossing's linewidth and re-emergence of thermal transition peaks (see Appendices \ref{Phase_matching} and \ref{Intrinsic_loss} for quantitative analysis on the thermal photon number). This enhanced decoherence is dominant at the critical coupling condition $\kappa_b = \kappa_i$ as it scales roughly as $ \sim \eta \sinh^2 r$, indicating that only by strongly over-coupling the resonator to the squeezed drive can one practically overcome the issue, rather than simply increasing the bandwidth of the squeezed vacuum drive or the squeezing strength \cite{Munoz2020}. Furthermore, with higher thermal squeezed occupation, the system is in fact showing classical-like response in the spectrum rather than genuine anti-crossing from vacuum Rabi coupling. These results agree with the conclusion from ref. \cite{Munoz2020}.

In Fig. (\ref{fig:intrinsic_loss}b) and (\ref{fig:intrinsic_loss}c), we plot the dynamics of the (lab-frame) photon number and the qubit, and we observe the signature of cavity QED system whose decoherence is dominated by thermal noise induced by the intrinsic losses. At critical coupling $\kappa_i / \kappa_b$, vacuum Rabi oscillation is completely destroyed and replaced by cavity thermalization towards a squeezed thermal state, a quality shared by its corresponding anti-crossing spectrum which only exhibits thermal peaks. However, at experimentally realizable outcoupling-to-loss ratio $\kappa_i / \kappa_b < 0.5$, we are still able to recover most of the oscillation. 

% Need to explore extra external squeezing requirement to counter intrinsic loss 

\section{Discussion}
\textit{Experimental feasibility:} While injecting broadband squeezed vacuum is frequently mentioned as a method to achieve effective strong coupling for systems driven by intracavity squeezing via a below-threshold OPO, the required bandwidth to reach the desired synthetically strong coupling from a weakly coupled system should be very large, especially considering the typical requisite squeezing values of $\gtrsim 15$dB. Since a cavity OPO threshold scales with the inverse square of its quality factor, this squeezing-bandwidth value becomes less practical. A viable alternative is using parametric waveguides - such as periodically-poled lithium niobate waveguide \cite{Nehra2022, Chen2022, Presutti2024} with THz optical bandwidths and Josephson travelling wave parametric amplifier \cite{Qiu2023} with GHz microwave bandwidths - as their bandwidths are much larger than that of an optical cavity (typically GHz to tens of MHz) or a microwave cavity (typically tens of kHz to sub-kHz). In addition, parametric waveguides do not have a threshold and therefore enable a path towards using broadband squeezed vacuum for enhancing light-matter interactions with strong continuous wave or pulse pumps. Furthermore, we note that in nonlinear nanophotonic waveguide, the squeezing bandwidth can simply be controlled by tuning the pump spectrum - e.g. when using ultrashort laser pulses \cite{Nehra2022}, or the waveguide poling, or both \cite{Hurvitz2023}, enabling experiments with tuneable squeezing bandwidth. 

\textit{Experimental implementation:} For optical cavity QED, a potential platform to realize this enhanced light-matter interaction is III-V semiconductors with self-assembled quantum dots. Specifically, InAs quantum dots have been shown to have single-photon coupling up to 25GHz in a gallium arsenide (GaAs) photonic crystal resonator (Fig. \ref{fig:proposal}(a)), with relatively narrow optical linewidth ($\gamma \sim 0.1 \textrm{GHz})$ \cite{Faraon2008, Muller2015}. While in these earlier experiments cavity $Q$ factor has been around 30,000 at 920nm quantum dot wavelength, there is room to improve this $Q$ by at least an order of magnitude, as limited by the material absorption loss. Furthermore, GaAs has both large $\chi^{(2)}$ and $\chi^{(3)}$ values \cite{Buckley2014}, enabling the generation of strong squeezing inside a cavity (when coupled to a squeezed bath) in a monolithic platform with low pump powers. A source of broadband squeezed vacuum in GaAs (or AlGaAs) could be realized by using a quasi-phase-matched poled waveguide that facilitates degenerate squeezing when pumped at the second-harmonic mode \cite{Skauli2002, Baboux2023}. Using a train of ultrafast pulses with tens of fs pulse-width, the squeezed vacuum could attain THz bandwidth, and a waveguide can be overcoupled to the photonic crystal cavity to ensure $\kappa > \kappa_i$ (Fig. \ref{fig:proposal}(a)). For an intrinsic $Q$ factor of $10^5$, the loaded $Q$ factor can be as low as $10^4$ while retaining $g/\kappa \sim 1$. The intracavity squeezing can then be realized via pumping at two frequencies $\omega_+, \omega_-$ such that squeezing can be generated via GaAs's $\chi^{(3)}$ (Fig. \ref{fig:proposal}(b)), all of which can be performed through free space optics to take advantage of photonic crystal's out-of-plane emissions, and therefore removes the need to have another waveguide to carry pump optical signals. We emphasize that for experimental implementation, it is best to work in the low detuning regime where $\delta_a, \delta_c \gtrsim g$ so that the parametric pump powers needed for realizing squeezing is within reasonable power range ($\mu \textrm{W} - \textrm{mW}$) for integrated platforms. To probe the enhanced coupling experimentally, we consider reading out the qubit's or the cavity's vacuum Rabi oscillation, which is accessible through through photonic crystal's out-of-plane emission. We show the corresponding simulation of the qubit's Rabi oscillation for squeezing on and no squeezing in Fig. \ref{fig:proposal}(c), assuming that the qubit is initialized in the excited state $|e\rangle$. The vacuum Rabi frequency of the system does indeed increase as compared to the original, un-enhanced Jaynes-Cumming model, albeit with similar increases in decoherence due to coupling to the intrinsic loss channel. Finally, we add that aside from III-V quantum dots, other platforms that support both optical nonlinearity and optically active quantum defects could also be viable for realizing our proposal such as SiC and diamond with lithium niobate. For example, SiC has large (but much lower than GaAs) bulk $\chi^{(2)}$ and $\chi^{(3)}$ nonlinearity, low optical loss and a diverse family of optically active color centers, and has experimentally demonstrated combined nonlinear optics with color centers in the same device \cite{Lukin2024}. Meanwhile, diamond's diverse color centers and ability to be heterogenously integrated with lithium niobate can take advantage of lithium niobate's large $\chi^{(2)}$ and mature fabrication of low-loss quasi-phase-matched waveguides or resonators \cite{Riedel2023}.

\section{Conclusion}

We conclude that injecting broadband squeezed vacuum is necessary for enhancing cavity QED. However, the bandwidth requirement far exceeds the capability of cavity-based squeezed light sources but are within reach for travelling wave or waveguide-based squeezed light source, or alternatively, intracavity squeezing using a strongly dissipative mode. We also show that intrinsic loss remains the major limiting factor for efficiently enhancing cavity QED with parametric drive at strong squeezing, and further reduction in loss is necessary to enable experiments with high squeezing for pushing the interaction strength well into the strong coupling regime. Together, we believe this approach will enable a better understanding for future experimental implementations in both integrated cavity QED with solid-state atomic defects and circuit QED with superconducting qubits to exploit injected and intracavity squeezed light towards quantum technology applications such as squeezed lasing \cite{Munoz2020}, fast entangling gates \cite{Ge2019, Burd2021, Burd2024}, fast quantum non-demolition qubit readout \cite{Qin2024}, enhancing weakly nonlinear bosonic system \cite{Lu2015, Lemonde2016, Yanagimoto2020}, and squeezed reservoir engineering for exotic macroscopic quantum systems \cite{Groszkowski2022, Karnieli2024}. 

\begin{acknowledgments}

This work has been supported by the Vannevar Bush faculty Fellowship from the US Department of Defense. Some of the computing for this project was performed on the Sherlock cluster. We would like to thank Stanford University and the Stanford Research Computing Center for providing computational resources and support that contributed to these research results. T.K.L acknowledges supports from NSF Graduate Research Fellowship. C.~R.-C. is supported by a Stanford Science Fellowship. A.K. is supported by the VATAT-Quantum fellowship by the Israel Council for Higher Education; the Urbanek-Chodorow postdoctoral fellowship by the Department of Applied Physics at Stanford University; the Zuckerman STEM leadership postdoctoral program; and the Viterbi fellowship by the Technion.

\end{acknowledgments}

\pagebreak

\clearpage

\appendix

\section{SLH derivation of simulation model}
In general, to account for driving a quantum system with output from another quantum system, we use the cascaded source model prescribed by SLH formalism, which models scattering, loss and Hamiltonian of a composite quantum system. Consider the SLH triples $G_1$ and $G_2$ of the source OPO and an arbitrary quantum system, respectively
\begin{align}
    G_1 &= (S_1, L_1, H_1) = (I, \sqrt{\kappa_a} \hat{a}, i(E_a\hat{a}^{\dagger 2} - E_a^* \hat{a}^2) \\ 
    G_2 &= (S_2, L_2, H_2) 
\end{align}
To depict the cascade between $H_1$ and $H_2$, we use the concatenation product rule, which is given by 
\begin{widetext}
    \begin{align}
    &G = G_2 \triangleleft G_1 = \left( S_2 S_1, L_2 + S_2 L_1, H_1 + H_2 + \frac{1}{2i} (L_2^\dagger S_2 L_1 - L_1^\dagger S_2^\dagger L_2) \right)
    \end{align}
\end{widetext}
For systems coupled via a common waveguide, $S_1 = S_2 = I$. The total SLH triple is 
\begin{align}
    G = \left(I, L_1 + L_2, H_1 + H_2 + \frac{1}{2i}(L_2^\dagger L_1 - L_1^\dagger L_2) \right) \label{eq:SLHTriple}
\end{align}
When applied to our system, we have 
\begin{align}
    L_1 + L_2 &= \sqrt{\kappa_a} \hat{a} + \sqrt{\kappa_b} \hat{b} \label{eq:SLH_jump} \\ 
    \frac{1}{2i}(L_2^\dagger L_1 - L_1^\dagger L_2) &= \frac{\sqrt{\kappa_a \kappa_b}}{2i} \left[ \hat{b}^\dag \hat{a} - \hat{a}^\dag \hat{b} \right]
    \label{eq:SLH_coup}
\end{align}
Eqs. (\ref{eq:SLH_jump}) and (\ref{eq:SLH_coup}) represent the outcouplings and couplings mediated by the common waveguide between the $\hat{a}$ and $\hat{b}$ modes, respectively. 

\section{Phase-matching and squeezing-matching conditions for squeezed bath} \label{Phase_matching}

\begin{figure*}[t]
    \centering
    \includegraphics[width = 18cm]{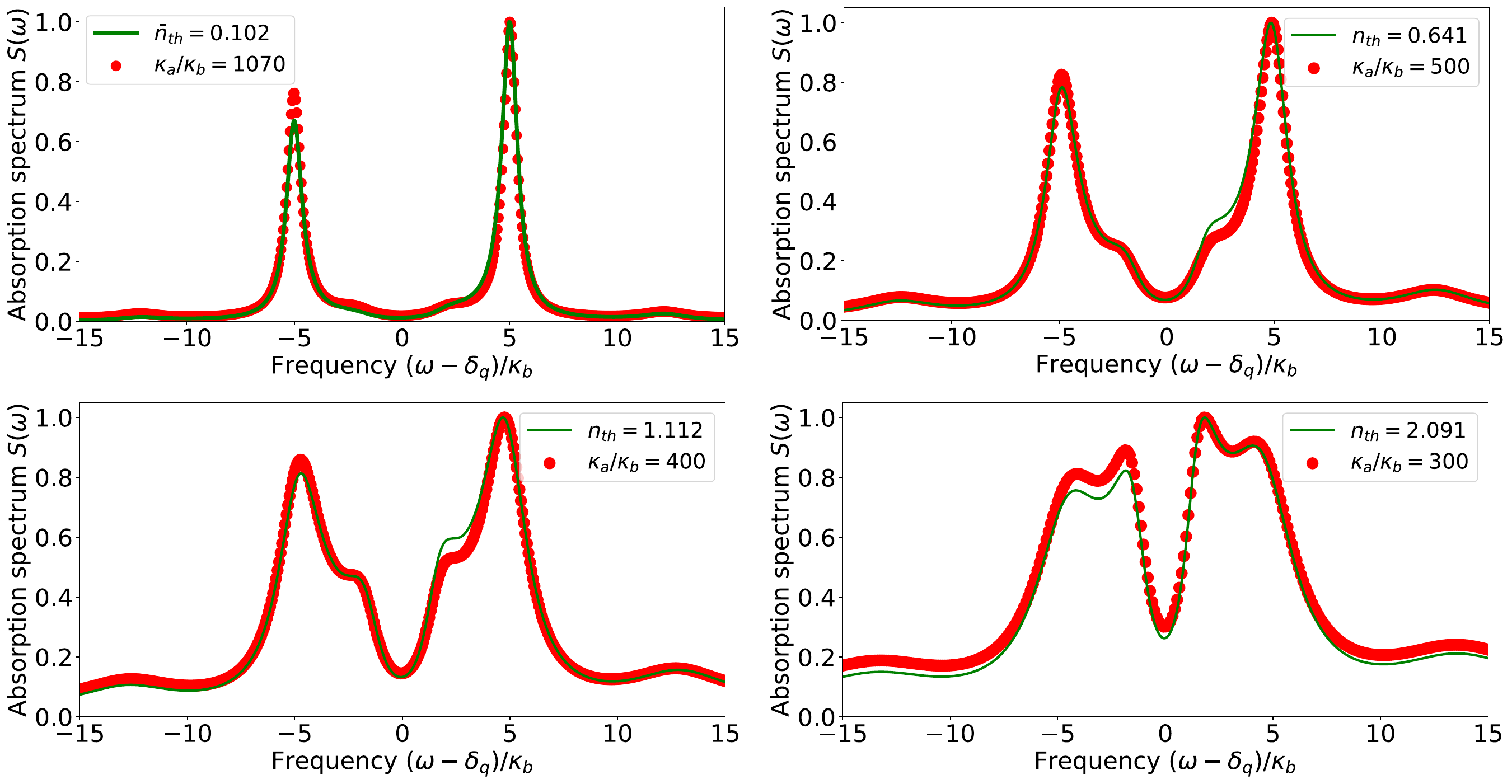}
    \caption{Comparison of best-fit absorption spectra using spectra computed from simulating Eq. \ref{eq:fit_me} that includes thermal and squeezing noises, and SLH absorption spectra. $n_\textrm{th}$ is obtained from computing $\lambda \sinh^2 r$. }
    \label{fig:fit}
\end{figure*}

To ensure that the Bogoliubov mode $\hat{\beta}$ is in the ground state, one can phase-match the phase and strength between the squeezed bath and intracavity squeezing. Here, we re-derive the matching conditions in ref \cite{Munoz2020}. In the lab frame, we can write the master equations as 
\begin{align}
    \frac{d\rho}{dt} = -i[H, \rho] + \frac{\kappa}{2} (N + 1 + \eta) L_{\hat{b}} [\rho] + \frac{\kappa}{2} N L_{\hat{b}^\dag} [\rho] + \notag \\ 
    - \frac{\kappa}{2} M D_{\hat{b}} [\rho] - \frac{\kappa}{2} M^* D_{\hat{b}^\dag} [\rho] 
\end{align}
Where $L$ and $D$ are the dissipators 
\begin{align}
    L_{\hat{b}}[\rho] &= 2 \hat{b}^\dag \rho \hat{b} - \hat{b}^\dag \hat{b} \rho - \rho \hat{b}^\dag \hat{b} \\ 
    D_{\hat{b}}[\rho] &= 2 \hat{b} \rho \hat{b} - \hat{b} \hat{b} \rho - \rho \hat{b}\hat{b}
\end{align}
With $N$ and $M$ are the squeezing coefficients 
\begin{align}
    N_s &= \sinh^2 r_e \equiv s_e^2  \\ 
    M_s &= \cosh r_e \sinh r e^{-i\theta_e} \equiv c_e s_e e^{-i \theta_e}
\end{align}
Where $r_e$ is the external squeezing coefficient. Now consider an injected squeezed bath. Performing the Bogoliubov transformation $\hat{\beta} = \cosh r \hat{b} - e^{-i\theta} \sinh r \hat{b} \equiv c \hat{b} - e^{-i \theta} s \hat{b} $, we have the new master equation 
\begin{align}
    \frac{d\rho}{dt} = -i[H, \rho] + \frac{\kappa}{2} (\mathcal{N}_s + 1 + \eta) L_{\hat{\alpha}} [\rho] + \frac{\kappa}{2} \mathcal{N}_s L_{\hat{\alpha}^\dag} [\rho] + \notag \\ 
    - \frac{\kappa}{2} \mathcal{M}_s D_{\hat{\alpha}} [\rho] - \frac{\kappa}{2} \mathcal{M}_s^* D_{\hat{\alpha}^\dag} [\rho]
\end{align}
Where the new squeezing coefficients are 
\begin{align}
    \mathcal{N}_s &= s^2( 1 + \eta) + N_s(c^2 + s^2) + cs(M_s e^{-i\theta} + M^*_s e^{i\theta})  \\ 
    \mathcal{M}_s &= cs e^{i\theta} (2N_s + 1 + \eta) + M_s c^2 + M^*_s e^{2i\theta} s^2 
\end{align}
To derive the phase matching conditions, we want $\mathcal{N}_s = \mathcal{M}_s = 0$ to suppress exponentially large decoherences $\kappa \sinh^2r, \kappa \cosh r \sinh r$. For the case where $\eta = 0$, we have 
\begin{align}
    \mathcal{N}_s &= s^2 + s_e^2 (c^2 + s^2) + cs c_e s_e (e^{-i(\theta + \theta_e)} + e^{i(\theta + \theta_e)}) \\ 
    \mathcal{M}_s &= cs e^{i\theta} (2s_e^2 + 1) + c_e s_e e^{-i \theta_e} c^2 + c_e s_e e^{i \theta_e} e^{2i\theta} s^2 
\end{align}
For $\mathcal{N}_s = 0$, we want $s^2 = s_e^2, c^2 = c_e^2$, which means $r = r_e$. This simplifies to 
\begin{align}
    \mathcal{N}_s = s^2 + s^2 (c^2 + s^2) + c^2 s^2 (e^{-i(\theta + \theta_e)} + e^{i(\theta + \theta_e)}) \label{eq:thermal_noise}
\end{align}
Finally, let $\theta + \theta_e = \pi$, we have $\mathcal{N}_s = s^2 + s^4 - c^2 s^2 = 0$. Applying these conditions on $\mathcal{M}_s$, we also have 
\begin{align}
    \mathcal{M}_s = e^{i\theta} \left[ cs(2s^2 + 1) - c^3 s - c s^3 \right] = 0
\end{align}
Thus, we have derived the necessary conditions on the squeezed bath such that we can exactly cancel out excessive noises from squeezing and recover Eq. (\ref{eq:sqz_bath_bogo}). When intrinsic loss $\eta$ is included, we show that it is still possible to cancel out $\mathcal{M}_s$ by solving the following relation 
\begin{align}
    \cos \theta [cs ( 2 s_e^2 + 1 + \eta ) - c_e s_e c^2 - s^2 c_e s_e ] = 0 
\end{align}
Equating this expression to zero, we have 
\begin{align}
    \sinh[2 (r - r_e)] + \sinh (2 r)  \eta = 0
\end{align}
So the external squeezing strength will be 
\begin{align}
    r_e = r + \frac{1}{2} \sinh^{-1}[ \eta \sinh (2 r) ]
\end{align}
Plugging $r_e$ back into Eq. (\ref{eq:thermal_noise}), we find 
\begin{align}
    \mathcal{N}_s = \frac{1}{2} \left[ 2 \eta \sinh^2 r + \sqrt{1 + \eta^2 \sinh^2 (2r)} - 1 \right] \label{eq:sqz_photon_num}
\end{align}
Which gives the following master equation % would be nice to add some simulations for this master equation
\begin{align}
    \frac{d \Tilde{\rho}}{dt} = -i[H, \rho] + \frac{\kappa}{2} (\mathcal{N}_s + 1 + \eta) L_{\hat{\beta}}[\rho] + \frac{\kappa}{2} \mathcal{N}_s L_{\hat{\beta}^\dag}[\rho] \label{eq:matched_int_loss}
\end{align}
We conclude that even with phase-matching and squeezing-matching conditions, intrinsic loss still poses a fundamental challenge to achieving strong coupling when both loss $\eta \gtrsim 0.1$ and squeezing value $r \gtrsim 1$ to keep $\mathcal{N}_s$ small, as the dissipator takes the form of thermal noises with strength $\mathcal{N}_s \kappa$.

% Informal titles, subject to changes 
% Don't read this part for now ... 
\section{Phase and squeezing strength matching in SLH}

%Redo this using adiabatic elimination argument 

To find the squeezing and phase-matching conditions in finite-bandwidth "squeezed bath", we adapt ref. \cite{Gardiner2004} derivation of output squeezed vacuum from a source OPO. We start with a source OPO in mode $\hat{a}$, which has output mode $\hat{a}_\textrm{out}(t)$ whose correlation functions are 
\begin{align}
     \langle \hat{a}_\textrm{out}^\dag(t) \hat{a}_\textrm{out}(t') \rangle &= \frac{\lambda^2 - \mu^2}{8} \left[ \frac{e^{-\mu(t - t')}}{\mu} - \frac{e^{-\lambda(t - t')}}{\lambda} \right] \\ 
     \langle \hat{a}_\textrm{out}(t) \hat{a}_\textrm{out}(t') \rangle &= \frac{\lambda^2 - \mu^2}{8} \left[ \frac{e^{-\mu(t - t')}}{\mu} + \frac{e^{-\lambda(t - t')}}{\lambda} \right]
\end{align}
Where 
\begin{align}
    \lambda &= \frac{\kappa_a}{2} + |E_a| \\
    \mu &= \frac{\kappa_a}{2} - |E_a|
\end{align}
Since $\hat{b}_\textrm{in}(t) = \hat{a}_\textrm{out}(t)$, the correlation functions for $\hat{b}_\textrm{in}(t)$ are the same as $\hat{a}_\textrm{out}(t)$. In the limit of large $\mu, \lambda$, the correlations can be approximated as Markovian, with the following coefficients
\begin{align}
    \langle \hat{b}_\textrm{in}^\dag(t) \hat{b}_\textrm{in}(t') \rangle = \langle \hat{a}_\textrm{out}^\dag(t) \hat{a}_\textrm{out}(t') \rangle &= N_s \delta(t - t') \\
    \langle \hat{b}_\textrm{in}(t) \hat{b}_\textrm{in}(t') \rangle = \langle \hat{a}_\textrm{out}(t) \hat{a}_\textrm{out}(t') \rangle &= M_s \delta(t - t')
\end{align}
Here 
\begin{align}
    N_s &= \frac{(\lambda^2 - \mu^2)^2}{4 \mu^2 \lambda^2} \\
    M_s &= \frac{\lambda^4 - \mu^4}{4 \mu^2 \lambda^2}
\end{align}
Finally, the phase of $M$, according to ref. \cite{Gardiner2004}, is $M = |M| e^{2i\phi}$, where $\phi = \textrm{arg}(E_a)$. In our simulation, we set $\phi = 0$, so according to phase-matching condition $\theta + \theta_e = \theta + 2 \phi = \pi$, we have $\theta = \pi$.  

% To analytically derive the phase matching condition, we adiabatically eliminate the mode $\hat{a}$ based on the SLH-ABCD formalism. Start with the SLH triple eq.(\ref{eq:SLHTriple}), we write the SLH Hamiltonian and the loss operator in the form 
% \begin{align}
%     -iH - \frac{1}{2} \sum_m L^\dagger L = k^2 Y^\dagger + k A^\dagger + B^\dagger
% \end{align}
% Assuming that $\kappa_a, \epsilon$ have the same scaling $O(k^2)$, we have 
% \begin{align}
%     k^2 Y^\dagger &= -iH_a - \frac{\kappa_a}{2} \hat{a}^\dagger \hat{a} \\
%     k A^\dagger &= \sqrt{\kappa_a \kappa_b} \hat{a} \hat{b}^\dagger \\ 
%     B^\dagger &= -iH_b - \frac{\kappa_b}{2} \hat{b}^\dagger \hat{b}
% \end{align}
% To find a "slow subspace" projector $P_0$, we project onto the squeezed vacuum of $\hat{a}$
% \begin{align}
%     P_0 = S_{\hat{a}} | 0 \rangle \langle 0 | S^\dagger_{\hat{a}} \otimes I_{\hat{b}}
% \end{align}
% Where $\hat{S}_{\hat{a}}$ is the squeezing operator of $\hat{a}$ with squeezing strength $r$ determined by 
% \begin{align}
%     r = \frac{1}{4} \ln \frac{1 + \epsilon / \kappa_a}{1 - \epsilon / \kappa_a}
% \end{align}
% Using this projector, we have a new SLH triple 
% \begin{align}
%     S &= I \\ 
%     L &= \sqrt{N_s + 1} \hat{b} - e^{i\phi} \sqrt{N_s} \hat{b}^\dagger \\
%     H &\rightarrow H_b 
% \end{align}
% Where $N_s = \sinh^2 r$ is the expected photon number. This SLH triple represents the "squeezed reservoir" that arises when a system interacts with infinitely broadband squeezed vacuum. 

\section{Analysis of finite bandwidth effects}
% Make this section an analysis on effective thermal effects with Rabi model + squeezed reservoir 
In this section, we explore an effective model for explaining decoherence processes at various injected squeezed vacuum bandwidth. We revisit the squeezed reservoir model but add thermal and squeezed noise channels with varying strength, such that in the squeezed frame the master equation becomes 
\begin{align}
    \frac{d\rho}{dt} &= -i[H, \rho] + D[\sqrt{\lambda N + 1} \hat{\alpha} - e^{i\theta} \sqrt{\lambda N} \hat{\alpha}^\dagger] (\tilde{\rho}) 
    \label{eq:fit_me}
\end{align}
Or, in the expanded form 
\begin{align}
    \frac{d\rho}{dt} &= -i [H, \rho] + \frac{\kappa}{2}(\lambda N + 1)D_\alpha [\rho] + \frac{\kappa}{2} \lambda N D_{\alpha^\dagger} [\rho] \notag \\ &- \frac{\kappa}{2} \lambda M D'_\alpha [\rho] - \frac{\kappa}{2} \lambda M^* D'_{\alpha^\dagger} [\rho]
    \label{eq:vacbath}
\end{align}
Where $\lambda \in [0, 1]$ is a free parameter for fitting. At $\lambda = 0$, we recover the ideal squeezed bath case, and at $\lambda = 1$, we have the master equation without the squeezed bath to counteract squeezing-induced noises. Physically, these decoherence channels have dissipation rates that scales as $\sim \lambda \sinh^2 r$. In Fig. \ref{fig:fit}, we fit the spectra generated by Eq. (\ref{eq:fit_me}) for various ratios $\kappa_a / \kappa_b$ through a grid search of $\lambda$ that minimizes the difference between the fitted spectra and the spectra generated via SLH simulation. Notice that at larger ratio $\kappa_a / \kappa_b$, the fit and SLH spectra tend towards agreement for spectrum features and approximate peak heights, which might be attributed to the fact that using a master equation as in Eq. (\ref{eq:fit_me}) implicitly assumes a purely Markovian response from the bath, which breaks down when $\kappa_a$ is not much larger than all other system's spectral values. The various smaller peaks that lie between the vacuum Rabi peaks at $\kappa_a / \kappa_b \in \{ 300, 400, 500\}$ are higher-rung transition peaks, which arise due to (incoherent) excitations of higher-energy levels of the qubit-cavity ladder \cite{Fink2010} due to coupling to a thermal reservoir. For further discussions on a strongly or ultrastrongly coupled system in thermal reservoir, see \cite{Settineri2018}.

\section{Simulation details}

% Use heat map for figure (a)
\begin{figure*}[t!]
    \centering
    \includegraphics[width = 16cm]{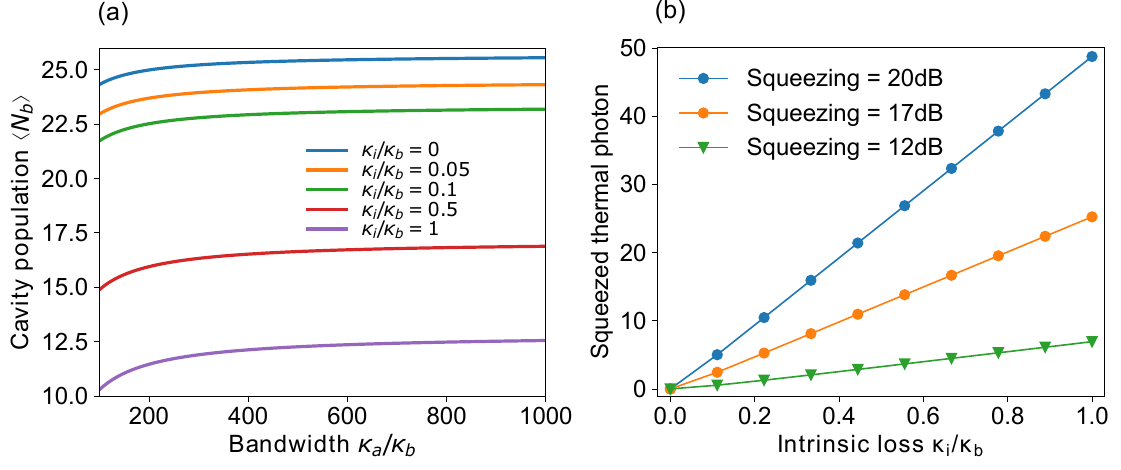}
    \caption{\textbf{(a)} Photon number in the second empty cavity driven by a finite-bandwidth squeezed vacuum source, as a function of bandwidth and intrinsic loss $\kappa_i$, with squeezing fixed at $\sim 20 \textrm{dB}$. \textbf{(b)} Effective squeezed thermal photon occupation $\mathcal{N}_s$ as a function of intrinsic loss and squeezing strength for the second cavity with squeezing drive (see Eq. \ref{eq:matched_int_loss})}
    \label{fig:loss_photon_num}
\end{figure*}

%Discuss how to derive the steady-state second cavity photon number, and discuss all parameters used in simulations. 

To keep the squeezing strength of the ideal squeezed reservoir case and the cascaded squeezing model the same, we compute the steady-state intracavity photon number $N$ \cite{Eddins2018} of the empty $\hat{b}$ cavity via input-output formalism \cite{Collet1984, Gardiner1985}
\begin{align}
    N_{b, ss} = \frac{2 \kappa_a |E_a|^2(2 \kappa_a + \kappa_b)}{\left[ \frac{\kappa_a^2}{4} - 4 |E_a|^2 \right] \left[ \frac{\kappa_a^2}{4} - 4 |\epsilon|^2 + \frac{\kappa_a \kappa_b}{2} + \frac{\kappa_b^2}{4} \right]} \label{eq:bss}
\end{align}
Here, we assume that $\kappa_i = 0$. Using Eq. (\ref{eq:bss}), we compute the necessary two-photon drive amplitude of the $\hat{b}$ cavity, $E_b = \textrm{tanh}^{-1}(r)$, where $r = \textrm{arcsinh}(\sqrt{N_{b, ss}})$, so that we can match the squeezing strength of the $\hat{b}$ mode with the $\hat{a}$ mode in simulation. 

% Discussion about Hilbert space truncation 
We note that for simultaneous injected external squeezing and intracavity squeezing in the SLH model, the Hilbert space truncation needs to be adjusted accordingly at higher squeezing to avoid accidentally throwing away higher Fock state occupation. Our simulations were performed in the Bogoliubov frame $\hat{\beta}$ as we can use a smaller number of Hilbert space truncation, resulting in faster simulation and convergence compared to lab frame Hamiltonian. We use $N_a = 25$ for all of our simulations, and we vary $N_b$ between $6$ and $16$ depends on the bandwidth $\kappa_a$, the squeezing strength $r$ and the effective squeezed thermal noise strength $\sinh^2 r$ - which is responsible for non-vacuum occupation in the Bogoliubov ladder \cite{Munoz2020, Villiers2024}. For the finite bandwidth regime, where $\kappa_a / \kappa_b \lesssim 400$, our simulations' accuracy is limited by the Hilbert space truncation. 

% \begin{figure}[h!]
%     \centering
%     \includegraphics{fig_1.eps}
%     \caption{Calibration of Fock state cut-off for quantum dynamics}
%     \label{fig:calibration}
% \end{figure}

For computing the anti-crossing spectra, we use the quantum regression theorem. In QuTiP \cite{Johansson2012, Johansson2013} implementation, we find the steady-state via direct method or LGMRES for large Hilbert space, then input into two-time correlation function to obtain $\langle \hat{\sigma}_-(t) \hat{\sigma}_+(0) \rangle $, and finally perform Fourier transform to obtain $S(\omega) $. To ensure smooth spectra, we sample from the correlation function for a time $t_f \gg 1/\gamma$. Finally, we note that computation time for obtaining spectra and quantum dynamics using parameters reported in this manuscript ranges between 20 hours to 48 hours, due to relatively slow convergence of QuTiP's master equation solvers with different parameter scales, as well as poorly conditioned Liouvillian matrices.

\section{Intrinsic loss model} \label{Intrinsic_loss}

In realistic experiments, a system will encounter intrinsic loss $\kappa_i > 0$ due to imperfection in device fabrications that cause scattering loss, material absorption and coupling to other decoherence channels such as another waveguide. Such a loss mechanism can be captured in the SLH model by adding $\kappa_i$ into the loss terms in the Lindbladian, and we assume that $\kappa_i$ is equal for both $\hat{a}$ and $\hat{b}$. With loss, the steady-state intracavity photon number $N$ of the cavity $\hat{b}$ without a two-level system and parametric drive is, according to input-output formalism 
\begin{widetext}
    \begin{align}
        N_{b, ss, \textrm{lossy}} = \frac{2 |E_a|^2 \kappa_a \kappa_b (2 \kappa_a + \kappa_b + 3 \kappa_i)}{ (\kappa_b + \kappa_i)\left[ \frac{(\kappa_a + \kappa_i)^2}{4} - 4|E_a|^2 \right] \left[ \frac{(\kappa_a + \kappa_b + 2\kappa_i)^2}{4} - 4|E_a|^2 \right]} 
        \label{eq:Nsi}
    \end{align}
\end{widetext}

Using Eq. (\ref{eq:Nsi}), we can solve for the expected squeezing value when $\kappa_i \neq 0$, thereby satisfying the squeezing strength matching condition. All other parameters in Eq. (\ref{eq:SLH_cavqed}) are kept the same. Note that Eq. (\ref{eq:Nsi}) reduces to Eq. (\ref{eq:bss}) when $\kappa_i = 0$.  In Fig. \ref{fig:loss_photon_num}(a) we show the steady-state cavity photon number fixed at 20dB as a function of bandwidth as well as intrinsic loss, and in Fig. \ref{fig:loss_photon_num}(b), the effective thermal occupation of the cavity in the squeezed frame as a function of intrinsic loss as well as in squeezing, according to Eq. (\ref{eq:sqz_photon_num}). To simulate the SLH master equation with intrinsic loss for Fig. \ref{fig:intrinsic_loss}, we work in the squeezed frame and truncate the Bogoliubov mode between 8 to 12 Fock states to account for higher thermal photon occupation in the squeezed Fock ladder. We stress that while at higher intrinsic loss, the squeezed frame shows that the number of squeezed thermal photon scales with $\kappa_i$, it is important to use the lab frame of the cavity to show that the intracavity state is in fact losing photons. 

\nocite{*}

\bibliography{apssamp}% Produces the bibliography via BibTeX.

\end{document}